\documentclass{article}
\PassOptionsToPackage{dvipsnames}{xcolor}
\usepackage{indexserve_arxiv,times}
\usepackage{booktabs,graphicx,microtype,xspace,xcolor}
\usepackage{amssymb,amsmath,amsfonts,bm}
\usepackage{subcaption,float,wrapfig,needspace,array}
\usepackage{hyperref,url}

\newcommand{\etc}{\hbox{\emph{etc.}}\xspace}
\newcommand{\sys}{\textsc{IndexServe}\xspace}
\newcommand{\ascendAThreeFull}{Huawei Ascend Atlas A3 SuperNode (CloudMatrix384)\xspace}
\newcommand{\ascendAThree}{Ascend A3\xspace}

\title{IndexServe: Disaggregating the Indexer as a\\Service for Efficient Sparse-Attention LLM Serving}
\author{\begin{tabular}{@{}c@{\quad}c@{\quad}c@{\quad}c@{}}
Yuzheng Wang$^{1,3}$ & Yilong Luo$^{1,3}$ & Junhao Hu$^{1,3}$ & Tiancheng Hu$^{1,3}$ \\
Wenbo Wu$^{1}$ & Wenpin Jiao$^{1}$ & Zhongzhe Hu$^{2}$ & Tao Xie$^{1,3}$ \\[0.6em]
\multicolumn{4}{@{}c}{\normalfont\small $^{1}$Peking University} \\
\multicolumn{4}{@{}c}{\normalfont\small $^{2}$Huawei Technologies Ltd.} \\
\multicolumn{4}{@{}c}{\normalfont\small $^{3}$Beijing Tongming Lake Information Technology Application Innovation Center}
\end{tabular}
}
\date{}

\begin{document}
\maketitle
\begin{abstract} 
Sparse attention reduces attention computation by attending to only a small subset of historical tokens. However, the indexer introduced to score all historical tokens still requires HBM storage that grows linearly with sequence length. In long-context serving scenarios, we find that the index state's growing memory demand exhausts HBM at low request concurrency, limiting the batch sizes and leading to low serving throughput.

We present \textsc{IndexServe}, an indexer-as-a-service system that disaggregates the indexer's HBM storage and selection computation from its original serving engine. Our key insight is that index selection requires no model weights, allowing a separate index service to pool index memory and computation across multiple inference instances. Disaggregating the indexer is challenging: remote selection lies on the critical path of every sparse-attention layer, while the shared service must handle dynamically growing indexes of varying lengths. To tackle these challenges, \textsc{IndexServe} leverages direct device-to-device communication and builds a graph-compatible runtime to minimize remote selection overhead, while introducing PagedIndex and length-aware dynamic scheduling to enable efficient index allocation and scheduling across the shared service.

As a result, \textsc{IndexServe} removes the last context-growing state from the serving engine's HBM, enabling larger batches and higher throughput. Evaluated on Ascend A3 with DeepSeek-V3.2 and GLM-5.1, \textsc{IndexServe}'s peak output throughput is 2.46--10.35$\times$ that of vLLM, SGLang, and ECHO.
\end{abstract}

 \section{Introduction}
\label{sec:introduction}
Large language models (LLMs) increasingly handle workloads such as long-document understanding, large code repositories, and agentic tasks, with contexts reaching hundreds of thousands or millions of tokens~\citep{longbenchv2,glm5}. Standard attention reads the key--value (KV) states of all previous tokens and computes attention over them, so both memory traffic and computation grow with context length. Sparse attention reduces this cost by letting each query attend to only a small subset of past tokens. Earlier methods typically introduce sparsity at inference time without changing model training~\citep{quest,infinigen,retrievalattention}. Recent models instead learn selection during training. DeepSeek Sparse Attention (DSA), adopted by recent large models from DeepSeek, GLM, \etc~\citep{deepseekv32,glm5,deepseekv4}, uses a lightning indexer to score all past tokens and select only the top-$k$ tokens, with $k=2048$. As the indexer is learned as part of the model, serving systems must preserve its selection rule.

The lower attention cost of sparse attention cannot fully translate this algorithmic saving into serving system's throughput gains.
A DSA model still keeps the full KV cache and additionally maintains a \emph{K index} (referred to as the \emph{index} below), which stores one compact index key for each past token and is scanned to select the top-$k$ tokens~\citep{deepseekv32}. Both states grow linearly with context length. 
As any past token may be selected at a later step, its KV and index must remain in the accelerator's high-bandwidth memory (HBM), which limits the number of requests that can run concurrently and thus caps the batch size the accelerator can reach, avoiding full use of its compute resources. This problem is especially severe during decoding: each request produces only one token per step, so efficient execution relies on batching many requests to amortize weight reads and kernel overhead~\citep{vllm,splitwise,distserve,sarathi}. Long-context state therefore creates a \emph{capacity wall} for the decoding stage in LLM serving: HBM limits the batch size before the sparse operators can reach high utilization.

To relieve this capacity pressure, serving systems such as SGLang and vLLM move KV out of HBM. The state-of-the-art approach moves the full KV cache from HBM to local host DRAM, referred to as \emph{KV offloading}~\citep{hisparsepaper,vllmsfa}. Each request keeps only a fixed-size \emph{sparse-KV buffer} in HBM for selected entries, fetching misses from DRAM on demand (Figure~\ref{fig:arch}). By removing the full KV cache from HBM, KV offloading allows more requests to run concurrently and increases the output throughput.

However, the other context-growing state in sparse attention, the index, must still remain fully in HBM because the indexer scans the complete history to select the top-$k$ tokens. Our measurements show that this index state can still make sparse-attention serving concurrency-bound and limit throughput. With DeepSeek-V3.2 at context lengths above 120K tokens, the current state-of-the-art KV offloading implementation~\cite{vllmsfa} admits only two concurrent requests per data-parallel (DP) replica, despite a configured limit of 16 max requests (Sec.~\ref{sec:compute}).
We characterize three root causes that make the index remain the HBM capacity wall. (Section~\ref{sec:capacity}). First, the index grows with context, while the sparse-KV buffer has a fixed size. At long contexts, the index takes most of the per-request HBM left after offloading. Second, current engines split HBM between the sparse-KV buffer and the index with a static partition set before requests arrive. No single split fits all context lengths, so part of HBM stays unused. Third, tensor parallelism (TP) cannot shard the index. All attention heads share one index key per token, so every TP rank stores a full copy and repeats the scan.

We present \sys, a shared indexer service for sparse-attention serving. \sys decouples the index state and scan/select compute from the decode stage and places them in a shared pool of accelerators (Figure~\ref{fig:arch}b). This separation is practical because scan/select needs no model weights: it uses only the decoder-produced index inputs and the stored index, so the historical index can remain in the service itself without model parameters. The decode instances (referred to as \emph{Decoders}) in Prefill-Decode disaggregation~\cite{distserve} keep the model weights and the remaining decode operators, while \sys stores the index and performs token selection. \sys follows the model's original scoring rule, numerical format, and top-$k$ rule, so it introduces no approximation to token selection. \sys is independent of KV offloading: KV offloading removes the KV cache from decoder HBM, while \sys removes the index. We build and evaluate \sys on top of KV offloading as it is the state-of-the-art for
sparse-attention serving, and together the two remove all of the growing states
from decoder HBM, unlocking the high-throughput potential of sparse attention.

\sys is designed to tackle two challenges. First, remote indexer sits on the critical
path of every sparse-attention layer, so it runs tens of times per decoding
step. Host-launched calls and CPU copies would break the kernel graph execution~\citep{cudagraphs} and add their overhead at every
layer. \sys therefore keeps remote selection inside graph replay, which runs a
recorded sequence of device operations without launching each one from the
host. It registers request and response buffers at fixed addresses before
capture and moves data directly between device buffers over the fast scale-up network on a \emph{supernode}~\cite{cloudmatrix,imex}. Second, one service must serve many decoders whose requests have
different and growing context lengths. Reserving memory for each request's
maximum context wastes HBM, and indiscriminately batching pending indexer tasks together delays
the completion latency. We introduce \emph{PagedIndex}, which grows each request's index page by page and
gives decoders stable logical handles, so the service can share memory and
change batch members without rebuilding graphs. Moreover, \emph{length-aware
    batching} uses offline profiles to size batches from currently ready tasks
according to the latency objective and load. It permits mixed lengths while
prioritizing short scans without starving long ones.

We implement \sys on the Huawei CloudMatrix384 supernode (referred to as \ascendAThree)~\citep{cloudmatrix}
and compare it with SGLang HiSparse\cite{hisparse}, vLLM sparse offloading\cite{vllmsfa}, and ECHO\cite{echo} on
DeepSeek-V3.2 and GLM-5.1 at long contexts. \sys's peak generation throughput
is 2.46--10.35$\times$ that of the three baselines, vLLM, SGLang and ECHO.

\Needspace{6\baselineskip}
Our contributions are threefold. First, we identify the decoder-resident
index as a capacity bottleneck that limits sparse-attention batching after
KV offloading (Section~\ref{sec:motivation}). Second, we design \sys to move
index storage and selection out of decoders without changing the model's
selection rule (Section~\ref{sec:design}). Third, we evaluate \sys on two
models against three systems, quantifying resource-normalized peak throughput
gains and their latency cost (Section~\ref{sec:evaluation}).
 \section{Background and Related Work}
\label{sec:background}
\paragraph{Sparse attention and its indexer.}
Sparse attention lets each query attend to a small subset of key--value (KV)
entries. In DeepSeek Sparse Attention (DSA), the lightning \emph{indexer}
scores each causally visible token $s$ for query token $t$ with $H_I$ query
heads and one shared index key per token, then keeps the top-$k$ token
identities~\citep{deepseekv32}:
\begin{equation}
 I_{t,s}=\sum_{j=1}^{H_I}w^I_{t,j}\,\operatorname{ReLU}
 \!\left((\mathbf q^I_{t,j})^\top\mathbf k^I_s\right),
 \qquad \mathcal S_t=\operatorname{TopKIndices}_{s\leq t}(I_{t,s},k).
 \label{eq:dsa-score}
\end{equation}
The query vectors $\mathbf q^I_{t,j}$ and weights $w^I_{t,j}$ are learned
projections of token $t$; $\mathbf k^I_s$ is the compact index key for token $s$.
Attention then reads only the KV entries named by $\mathcal S_t$, and combines
values with softmax weights from scaled query--key dot products. These
attention vectors are separate from the indexer vectors.
The index keys of all visible tokens form the \emph{K index}, which we call the
\emph{index} from here on.
Attention touches $k$ tokens, but the index and its per-query scan in
Equation~\ref{eq:dsa-score} still grow with the visible context length $L$.
Sparsity comes in two forms. Training-free methods such as QUEST add selection
to a pretrained dense model~\citep{quest}.
Training-aware models learn to attend sparsely during training as in DSA deployed by DeepSeek-V3.2 and
GLM-5.1~\citep{nsa,deepseekv32,glm5}.

\paragraph{Algorithm-level optimizations change what is selected.}
One line of work lowers sparse-attention cost by rewriting selection itself.
HISA first scores pooled block representatives, then restricts token-level
selection to the kept blocks, which approximates DSA selections and needs no
extra training~\citep{hisa}. IndexCache reuses selected token identities across
neighboring layers to skip indexer calls~\citep{indexcache}.
GLM-5.2's IndexShare moves cross-layer reuse into training, so one indexer
supplies selections to a group of four layers~\citep{indexshare}.

\paragraph{System-level optimizations keep selection fixed.}
SGLang HiSparse, vLLM sparse offloading (Figure~\ref{fig:arch}a), and ECHO
preserve selection rules and numerics while offloading full KV to host DRAM
and retaining a fixed-size sparse-KV buffer of $h\geq k$ tokens per request
on the accelerator~\citep{hisparsepaper,vllmsfa,echo}.
HiSparse integrates cache management and transfers into the decode graph,
while ECHO overlaps transfers with indexer computation via lossless
prefetching. All three retain the index on the inference engine.

\paragraph{Our scope.}
\sys moves the index of training-aware sparse models such as DSA from decoder
HBM to other deployed accelerators, preserving the deployed model's scoring
expression, numerical format, and top-$k$ rule without adding selection
approximations.
It complements algorithm-level optimizations and can operate with or without
KV offloading; our evaluation uses KV offloading.
 \section{Motivation: The K Index Caps Decode Batching}
\label{sec:motivation}
A DSA request stores two states that grow with context length: the KV cache
and the index. KV offloading moves the KV cache to DRAM, so decoder HBM then
holds a fixed-size sparse-KV buffer and a growing index per request.
Section~\ref{sec:compute} shows that the decode batch stays small while the
sparse stages could still use larger batches. Section~\ref{sec:capacity} gives
three reasons why the index sets this limit, and
Section~\ref{sec:opportunity} states the opportunity. All results use the
DeepSeek-V3.2 INT8 deployment.

\subsection{Index Capacity Limits Decode Batching}
\label{sec:compute}
\begin{figure}[!htbp]
\centering
\begin{subfigure}[t]{0.24\linewidth}
\centering\includegraphics[width=\linewidth]{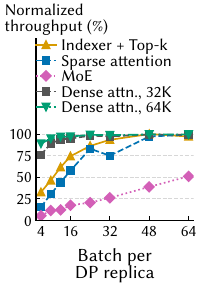}
\caption{}
\end{subfigure}\hfill
\begin{subfigure}[t]{0.24\linewidth}
\centering\includegraphics[width=\linewidth]{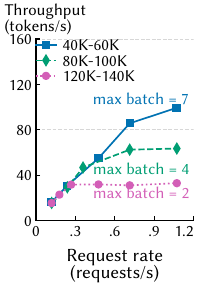}
\caption{}
\end{subfigure}\hfill
\begin{subfigure}[t]{0.24\linewidth}
\centering\includegraphics[width=\linewidth]{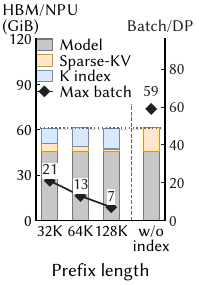}
\caption{}
\end{subfigure}\hfill
\begin{subfigure}[t]{0.24\linewidth}
\centering\includegraphics[width=\linewidth]{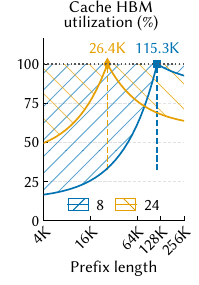}
\caption{}
\label{fig:partition-utilization}
\end{subfigure}
\caption{Batching limits in DeepSeek-V3.2 INT8.
(a) Stage throughput normalized to each stage's maximum.
(b) vLLM sparse offloading on Workloads A--C; labels show maximum active
batch per DP replica.
(c) Analytical HBM capacity with a BF16 index and per-length optimal
sparse-KV/index split; ``w/o index'' removes the decoder-resident index.
(d) Analytical static-partition HBM utilization at maximum concurrency
for \texttt{max-num-seqs} of 8 and 24; hatching marks unused HBM
and dashed lines mark full-utilization prefixes.}
\label{fig:memory}
\end{figure}

\paragraph{Sparse attention raises the batch needed to fill the hardware.}
Decoding is memory-bound at small batches, and batching spreads weight reads
over more requests~\citep{vllm,splitwise,distserve,sarathi}. Sparse attention
makes batching matter more: dense long-context attention is bandwidth-bound and
saturates early, whereas sparse attention reads only $k$ tokens per query, so
small batches leave it bound by launch and latency costs.
Figure~\ref{fig:memory}a measures this gap by profiling each stage alone over
batches 4--128: the indexer and sparse attention at a 64K
context, and dense attention at 32K and 64K.
At batch 8, dense attention reaches 89--95\% of its maximum throughput,
versus only 46\%, 31\%, and 12\% for the indexer, sparse attention, and MoE.
Larger batches improve all three; the indexer levels off near batch 48,
while sparse attention and MoE continue improving.

\paragraph{The measured batch stops far below the configured limit.}
Figure~\ref{fig:memory}b serves the three workloads of
Section~\ref{sec:eval-setup} with vLLM sparse offloading. Throughput flattens
early: as context length grows, the maximum active batch per DP replica reaches
only 7, 4, and 2, against a configured limit of 16. This panel counts admitted
requests in end-to-end serving at 40K--140K contexts. Together with
Figure~\ref{fig:memory}a, this result shows that the batch stays small while the
sparse stages would still gain throughput from more requests.

\paragraph{The HBM-resident index leaves substantial batching potential untapped.}
Figure~\ref{fig:memory}c shows analytical upper bounds on the maximum batch size. After model
weights, each NPU keeps about 15.9~GiB of HBM for the sparse-KV buffer and K
index. Even under an ideal split optimized separately for each prefix length, the bound falls
from 21 to 13 to 7 requests between 32K and 128K, because each request's index
grows while its sparse-KV buffer stays fixed. Without the index, the same
budget admits 59 requests at any prefix length, 2.8--8.4$\times$ more.

\Needspace{18\baselineskip}
\subsection{Why the Index Sets the Limit}
\label{sec:capacity}
\begin{wraptable}[16]{r}{0.49\linewidth}
\centering
{\footnotesize\setlength{\tabcolsep}{4pt}
\begin{tabular}{@{}rrrr@{}}
\toprule
Prefix & K index & \shortstack{DRAM full\\KV} &
\shortstack{Sparse-KV\\buffer} \\
\midrule
32K & 0.477\,GiB & \shortstack{2.145\,GiB\\(22.2\%)} & \shortstack{0.268\,GiB\\(1.78$\times$)} \\
\addlinespace[2pt]
64K & 0.953\,GiB & \shortstack{4.289\,GiB\\(22.2\%)} & \shortstack{0.268\,GiB\\(3.56$\times$)} \\
\addlinespace[2pt]
128K & 1.906\,GiB & \shortstack{8.578\,GiB\\(22.2\%)} & \shortstack{0.268\,GiB\\(7.11$\times$)} \\
\bottomrule
\end{tabular}
}
\caption{Per-request payload for DeepSeek-V3.2 across 61 layers
(BF16, 4K-token sparse-KV buffer). Parentheses give index/full-KV and
index/sparse-buffer ratios.}
\label{tab:request-memory}
\end{wraptable}
\paragraph{Index growth dominates the HBM capacity requirement.}
After KV offloading~\citep{vllmsfa}, the index is the only per-request state in decoder HBM
that still grows. In each DeepSeek-V3.2 layer, a context token contributes
1,152 bytes to the multi-head latent attention (MLA) KV cache and 256 bytes to
the index~\citep{deepseekconfig}, whereas each request reserves a fixed
4,096-token sparse-KV buffer~\citep{hisparse,vllmsfa}.
Table~\ref{tab:request-memory} quantifies how this imbalance grows with context length. Even at a
32K-token context, the index occupies 0.477~GiB per request, exceeding the
0.268~GiB sparse-KV buffer. At 128K tokens, the index grows to 1.906~GiB, 7.11$\times$ the sparse-KV buffer size, so the index accounts for most of the per-request
HBM requirement.

\paragraph{A static partition leaves HBM unused.}
Current serving engines statically partition HBM between the index and the
sparse-KV buffer before requests arrive for efficient memory management. Because their memory demands scale
differently with sequence length, no fixed split is optimal across lengths. SGLang
HiSparse sets index capacity as a configured multiple of sparse-KV
capacity~\citep{hisparse}; vLLM sparse offloading reserves sparse-KV buffers
for the configured concurrency and gives the rest of HBM to the index.
Figure~\ref{fig:memory}d shows what follows. At short prefixes,
part of the index pool sits idle; at long prefixes, the index pool fills first
and reserved sparse-KV slots go unused. A static partition therefore wastes
HBM on top of the index usage itself.

\paragraph{TP cannot shard the index.}
In DeepSeek-V3.2, MLA KV and the index each store one vector or key shared
across query heads, so neither can be sharded by KV head under tensor
parallelism (TP). Every TP rank therefore stores the same index keys and repeats
scan/select, which duplicates both HBM consumption and computation work. This replication makes the
HBM capacity wall that bounds the batchsize of decoding phase worse.

\subsection{Insight and Opportunity}
\label{sec:opportunity}
Three properties make moving the index out of the serving engine's HBM practical.

\textbf{Scan/select needs no model weights.}
Unlike projections, scan/select needs only the current token's index query,
new index key, and scoring coefficients, plus stored keys
(Equation~\ref{eq:dsa-score}), allowing the indexer to execute without model
parameters. Keeping historical keys there makes the payload in each direction
independent of context length, at about 8~KiB per layer for fixed $H_I$, $d_I$,
and $k$ (Appendix~\ref{app:communication}).

\textbf{Supernode scale-up fabrics provide fast, fine-grained communication.}
Modern supernode scale-up fabrics provide high bandwidth, low latency, and efficient fine-grained data access without traversing PCIe or RNIC
hops~\citep{cloudmatrix,xdeepserve,imex}. We use Huawei \ascendAThreeFull as a concrete
platform to demonstrate \sys's design (Appendix~\ref{app:hardware}) and discuss
its portability to NVIDIA platforms in Appendix~\ref{app:nvidia-portability}.

\textbf{A shared indexer service further improves cluster-wide utilization.}
The computation of indexer occupies only a short portion of each decode step, so an indexer
service dedicated to one decoder would sit idle while that decoder runs other
operators. A shared Indexer Service instead serves multiple decoder instances,
pooling both their index storage and computation, improving overall resource
utilization.
 \section{Design}
\label{sec:design}

\begin{figure}[!htbp]
\centering\includegraphics[width=0.9\linewidth]{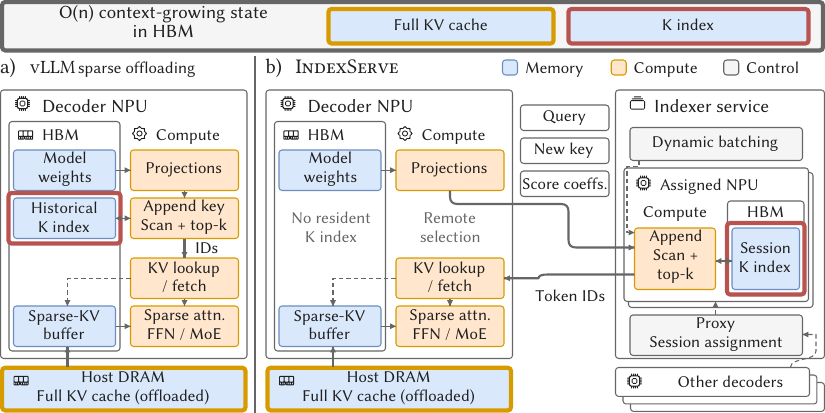}
\caption{Removing the two $O(n)$ HBM states ($n$: context length).
(a) vLLM offloads full KV to host DRAM but retains K index storage and
scan/select on the decoder. (b) \sys moves both to service NPUs.
Orange and red outlines mark full KV and K index storage, respectively;
dashed arrows denote control dependencies.}
\label{fig:arch}
\end{figure}

\subsection{Indexer as a Service}
\label{sec:service-overview}
As shown in Figure~\ref{fig:arch}b, \sys moves the index and scan/select into a dedicated accelerator pool (Section~\ref{sec:opportunity}), freeing decoder HBM for
larger batches. Decoders retain model weights, projections, sparse attention,
FFN/MoE, and KV storage, which contains a sparse-KV buffer in HBM and the full KV cache in host
DRAM. The indexer service stores and appends index keys, scans the contexts, and
returns top-$k$ token IDs. It uses the same scoring expression, numerical
format, and top-$k$ rule as the original serving engine, preserving both selection
results and numerical precision.

In indexer service, a proxy manages the indexes of sessions across multiple decoders. During
session setup, it selects a service accelerator based on free index capacity
and scan load, transfers the full context index once, and records the route. The
index remains on that selected accelerator throughout decoding, and subsequent indexer
calls from the session follow the same route. During steady-state decoding, only the current-step query, new index key, scoring coefficients, and selected IDs cross the decoder--service boundary, none of which grow in size with context length (Appendix~\ref{app:communication}).

\subsection{Graph Replay and Remote Communication}
\label{sec:graph}
\begin{figure}[t]
\centering\includegraphics[width=0.9\linewidth]{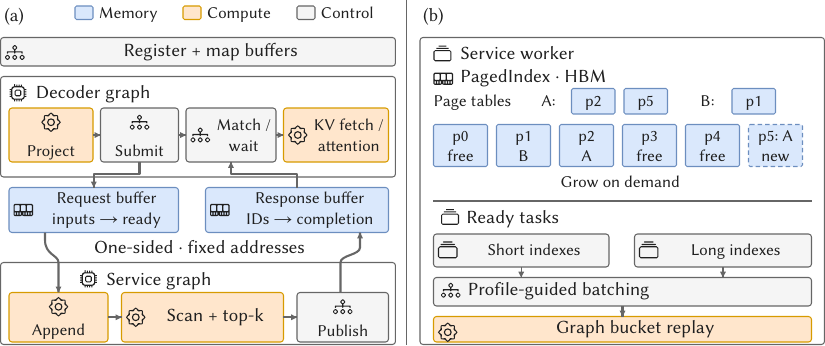}
\caption{Indexer Service mechanisms.
(a) Graph replay with fixed communication buffers and device-side
synchronization.
(b) PagedIndex with on-demand allocation and length-aware,
profile-guided batching.}
\label{fig:design}
\end{figure}

Remote selection sits on the critical path of every DSA layer, so
host-issued launches and CPU staging would repeat their overhead tens of times
per token. \sys therefore keeps remote selection inside
graph replay, which runs a recorded sequence of operations without launching
each one from the host (Figure~\ref{fig:design}a). Before capture, the runtime
registers request and response buffers and maps them for remote device access;
their addresses stay fixed across replays while their contents change with each
call.

After projection, device operations send the inputs over the supernode fabric
and wait for the matching result, moving tensor data directly between device
buffers. The service also replays captured graphs for its supported batch
sizes. Because each captured graph reads row descriptors from fixed buffers,
the scheduler changes which sessions occupy which rows by rewriting
descriptors, not by rebuilding the graph. On both paths,
data must be visible before the sender signals readiness or
completion.

\subsection{Index Memory and Scheduling}
\label{sec:service-management}
A shared indexer service holds indexes for sessions whose final lengths are unknown.
Reserving each session's maximum context would waste the pooled memory, and
handles tied to one engine's memory layout would make state shared by several
engines harder to manage.
\emph{PagedIndex} therefore grows each session's index as needed
(Figure~\ref{fig:design}b). It divides service HBM into fixed-size pages and
maps logical token positions to those pages. Appending a key fills the last
page or allocates another, without moving earlier keys. Decoders use stable
logical handles, and the service tracks page locations and frees pages after
their last use.

Scheduling pooled indexer requests is challenging. Naively batching all queued
tasks across decoders and sessions can exceed the batch size at which scan
throughput saturates and delay the completion of short-context indexes. We therefore profile scan throughput and execution time across index lengths
and supported batch sizes. At each dispatch, the scheduler uses currently ready
tasks and selects a batch size according to the latency objective and current
load. Additionally, Length-aware queues guide priority: short-index queues receive
more opportunities, while recurring long-index opportunities and oldest-first
selection prevent starvation. Admission holds back new sessions when memory or scan capacity is
exhausted. See more details in Appendix~\ref{app:design-details}.
 \section{Evaluation}
\label{sec:evaluation}

\subsection{Setup}
\label{sec:eval-setup}
We compare \sys with vLLM sparse offloading~\citep{vllmsfa},
SGLang HiSparse~\citep{hisparsepaper}, and ECHO~\citep{echo} on
\ascendAThree, using DeepSeek-V3.2 INT8 and GLM-5.1 INT4.
All baselines run on Ascend backends, including our reimplementation of ECHO's
early KV prefetching and indexer-kernel fusion in SGLang.
The main comparison uses 16 decoder NPUs (TP8$\times$DP2).
Workloads A--C replay the same InfiniteBench~\citep{infinitebench} requests
with 40K--60K, 80K--100K, and 120K--140K contexts and a 128-token output cap.
All systems keep full KV in DRAM, and we measure decoding only: each request
starts generation once its precomputed KV cache and index are loaded.
We report mean ITL and normalize \sys throughput by $16/(16+n)$, where $n$ is
the number of service NPUs used in that run. Idle NPUs on the dedicated service
server are left out of this count.
Appendix~\ref{app:eval-setup} gives evaluation setup details.

\subsection{Throughput}
\begin{figure}[t]
\centering\includegraphics[width=\linewidth]{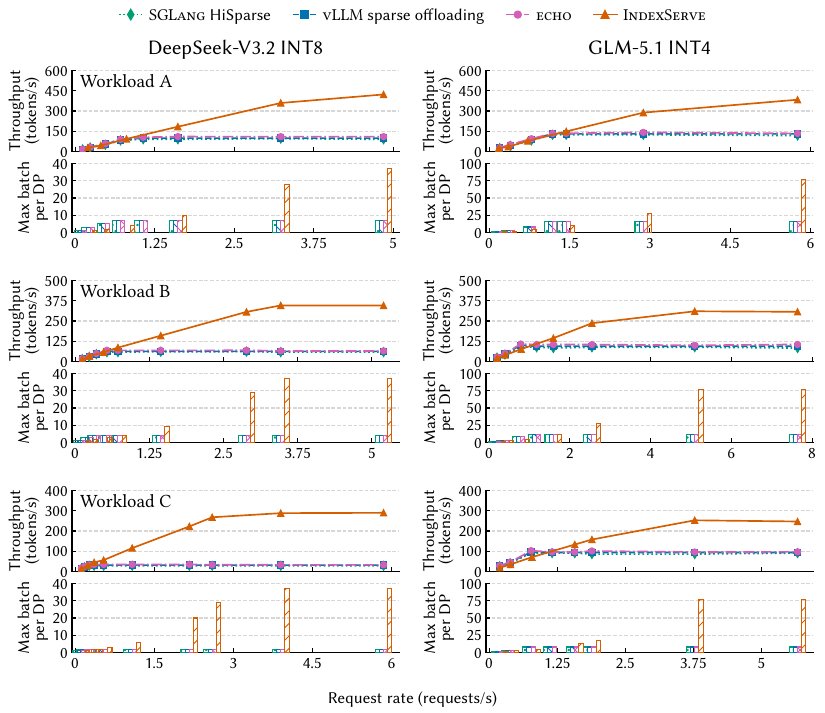}
\caption{Generation throughput (upper lines) and maximum batch per DP
replica (lower bars) for Workloads A--C, TP8$\times$DP2.
\sys throughput uses the normalization in
Section~\ref{sec:eval-setup}. Each pair shares a linear request-rate
 axis; bars have small system-specific offsets around measured rates.}
\label{fig:rate_throughput}
\end{figure}
Each ratio below compares two systems at their own peak throughput, reached at
different request rates and latencies, under the accounting of
Section~\ref{sec:eval-setup}.
Freeing index HBM lets more requests share decoder weight reads and kernel
launches (Section~\ref{sec:compute}).
In Figure~\ref{fig:rate_throughput}, DeepSeek's maximum batch stays
at 37 as contexts grow, while baseline batches fall from 7 to 2.
This batch gap matches its gain over ECHO growing from 3.82 to
8.32$\times$.
GLM's smaller weights leave the baselines more HBM, and its gain is
2.46--2.88$\times$. The benefit is therefore largest when index memory
limits the batch. ECHO's KV prefetching cuts transfer delay but keeps
this memory limit. Appendix~\ref{app:partition} tests other memory splits.

\begin{figure}[t]
\centering\includegraphics[width=\linewidth]{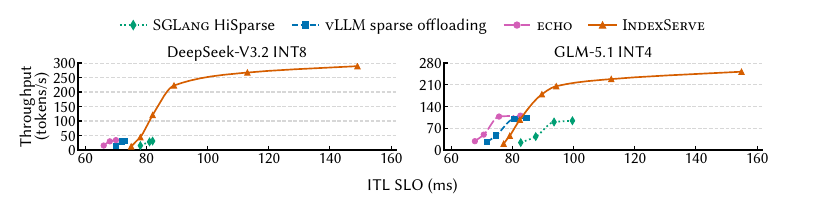}
\caption{SLO-constrained decoding throughput. Maximum generation throughput under varying mean-ITL budgets. \sys throughput uses the normalization in
Section~\ref{sec:eval-setup}.}
\label{fig:slo_throughput}
\end{figure}
\noindent\textbf{SLO-constrained decoding throughput.} Figure~\ref{fig:slo_throughput} reports the highest throughput each system
sustains while mean ITL stays within the Service Level Objective (SLO). Under stringent latency constraints (e.g., 80~ms for DeepSeek-V3.2 INT8), \sys offers limited benefits as remote selection introduces additional communication and synchronization overhead.
As the ITL budget is moderately relaxed (e.g., from 80 to 100~ms), \sys achieves substantial throughput improvements by admitting larger decode batches. In contrast, the colocated baselines quickly reach their throughput ceilings because the decoder-resident index limits batch concurrency, preventing them from effectively exploiting the additional latency budget.
As the SLO is relaxed further (e.g., for offline inference or batch processing scenarios), \sys's throughput gains gradually slow down and at a 150 ms mean-ITL budget, \sys achieves maximum throughput.

\subsection{Latency}\begin{figure}[t]
\centering\includegraphics[width=\linewidth]{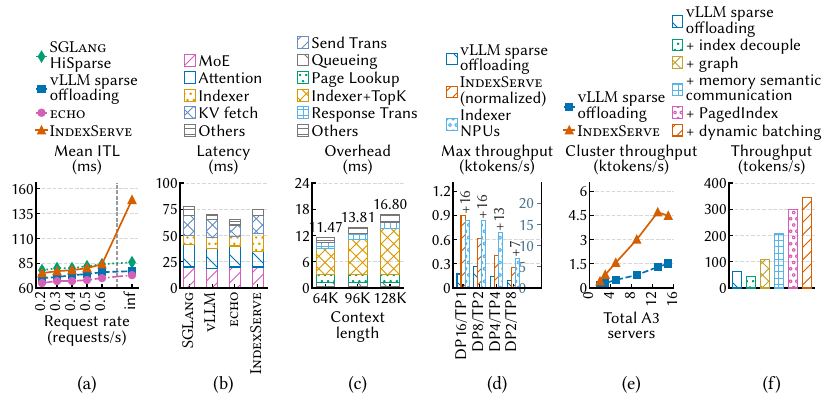}
\caption{Latency and throughput analysis.
(a)--(c) and (f): DeepSeek-V3.2 INT8;
(d)--(e): GLM-5.1 INT4, Workload C.
(a) Mean ITL versus request rate; \texttt{inf} marks each system's
peak-throughput point.
(b) Exposed ITL breakdown at 0.2 requests/s.
(c) Indexer Service overhead versus context length, local batch 12.
Times in (b,c) are summed across layers.
(d) Peak throughput across DP$\times$TP layouts on 16 decoder NPUs;
\sys is resource-normalized, with service NPU counts on the right axis.
(e) Unnormalized cluster throughput versus total A3 servers,
TP8$\times$DP2, with one \sys server dedicated to the service.
(f) Cumulative peak-throughput ablation, TP8$\times$DP2, Workload B;
mechanisms accumulate left to right.}
\label{fig:rate_latency}
\label{fig:in_depth}
\label{fig:ablation}
\end{figure}
Figure~\ref{fig:rate_latency}(a) shows how mean ITL changes as the request rate
increases. \sys maintains a mean ITL of 75--84~ms as the request rate increases
from 0.2 to 0.6 requests/s. Its mean ITL reaches 149~ms when the system operates
at peak throughput. Figure~\ref{fig:rate_latency}(b) breaks down the exposed ITL
at 0.2 requests/s. Remote selection adds data transfer, page lookup, and service
waiting to the critical path. Summed across all model layers in one decode step,
the indexer takes 16.8~ms for \sys, compared with 9--10~ms for the baselines.
ECHO
stays faster because it also exposes less KV-fetch time. Figure~\ref{fig:rate_latency}(c) further isolates the effect of context
length at a fixed batch size of 12. As the context length increases from 64K to
128K, the scan-and-select time grows from 6.0 to 10.6~ms, and the total Indexer
Service latency increases from 11.5 to 16.8~ms.

\subsection{In-Depth Analysis}
\label{sec:parallel-scale}
\noindent\textbf{Decoder parallelism.}
TP partitions model weights across NPUs but introduces index and KV-cache
HBM redundancy across TP ranks. DP avoids this cross-rank state redundancy but
replicates model weights across decoder instances. We evaluate four
DP$\times$TP layouts with a fixed budget of 16 decoder NPUs.
Figure~\ref{fig:rate_latency}(d) shows that \sys outperforms vLLM by
2.30--5.16$\times$. The 5.16$\times$ gain at TP1 shows that the benefit does not
depend on eliminating TP-induced index replication. The Indexer NPU labels
count assigned NPUs rather than fully utilized NPUs, as indexer workloads
are dispatched across NPUs for better load balancing.
Appendix~\ref{app:model-parallel} reports the complete request-rate sweeps
across all four layouts and workloads.

\noindent\textbf{Scalability.}
With one A3 server fixed for the Indexer Service, throughput scales
nearly linearly as we increase the number of decoder instances,
reaching 4,690 tokens/s with 12 decoder instances and 13 servers in total
(Figure~\ref{fig:rate_latency}(e)).
Adding more decoder instances causes throughput to drop by 4.7\%, as the index service reaches its HBM capacity limit,
forcing new requests to wait for available index memory.
Despite this limit, \sys still achieves 2.96$\times$ the throughput
of vLLM at 15 total servers.
Adding more index service servers is a possible way
to handle higher demand and we leave this implementation to future work.

\noindent\textbf{Ablation.}
We add the mechanisms of Section~\ref{sec:design} one at a time, in the order
we built \sys, and measure the peak throughput of each configuration
(Figure~\ref{fig:rate_latency}(f)).
Moving the index to the service alone is slower than the baseline, 45.2 versus
62.4 tokens/s. It admits a much larger batch, but host-issued calls and
per-layer transfers make each forward pass slower.
Graph replay and memory-semantic communication (Section~\ref{sec:graph}) cut
that per-layer cost, which lowers latency and raises throughput to 109.9 and
207.5 tokens/s.
PagedIndex then removes the service-side static index reservation that had
capped the batch. It reaches 299.2 tokens/s and needs fewer service NPUs.
Dynamic batching lowers each request's indexer latency
(Section~\ref{sec:service-management}) and reaches 346.1 tokens/s,
5.54$\times$ the baseline.

\noindent\textbf{Scope of the measurements.}
For the two models and three workloads in Section~\ref{sec:eval-setup}, we
report resource-normalized decoding output throughput.  \sys preserves the deployed
scoring rule, numerical format, and top-$k$ rule by design
(Section~\ref{sec:service-overview}). We further validate that its selected
token IDs and model quality match those of vLLM Ascend, with 100\% greedy
next-token agreement.
 \section{Conclusion}
\label{sec:conclusion}
Long-context sparse-attention serving becomes increasingly constrained as the
index grows with sequence length while remaining in decoder-local HBM. Its
growing HBM occupation reduces decode concurrency, limits batch size, and
prevents sparse attention from translating its computational savings into
serving throughput. We present \sys, which disaggregates index storage and
selection from decoders and provides them through a shared Indexer Service.
The service pools index capacity and computation across decoder instances,
preserves the original selection semantics, and allows decoder and indexer
resources to be provisioned independently. Our evaluation demonstrates that
this service-based disaggregation relieves the decoder HBM capacity bottleneck
and enables sparse-attention serving to sustain higher concurrency and
throughput at long context lengths.
 
\bibliography{references}
\bibliographystyle{plainnat}

\clearpage
\appendix
\section{Ascend A3 SuperNode, A3 Server, and 910C NPU}
\label{app:hardware}

This appendix gives the platform-specific implementation of the graph and
memory-semantic communication design in Section~\ref{sec:graph}.
The design needs mapped remote memory, device-initiated transfers, and
explicit completion ordering. The mechanisms below meet those requirements on
\ascendAThree; their names and topology are not part of the service interface.

\subsection{Hardware hierarchy and memory placement}
This paper uses three hardware terms at three levels. The \emph{Ascend A3
supernode} is the complete scale-up system. It spans 12 compute racks and four
communication racks and contains 48 \emph{A3 servers}, 768 Ascend 910C NPUs,
and 192 Kunpeng CPU sockets. An \emph{A3 server} is the unit that the public
CloudMatrix paper calls a compute node. It contains 16 \emph{910C NPUs}, four
CPU sockets, and seven first-level Unified Bus (UB) switches. We count servers
when describing deployment scale and NPUs when describing parallelism or
resource allocation. Huawei's English product documentation identifies the Atlas 900 A3
SuperNode as the A3 cluster basic unit~\citep{ascenda3products}. Its English
scheduling manual gives the corresponding multi-server
hierarchy~\citep{ascenda3affinity}.

\begin{figure}[ht]
\centering
\includegraphics[width=\linewidth]{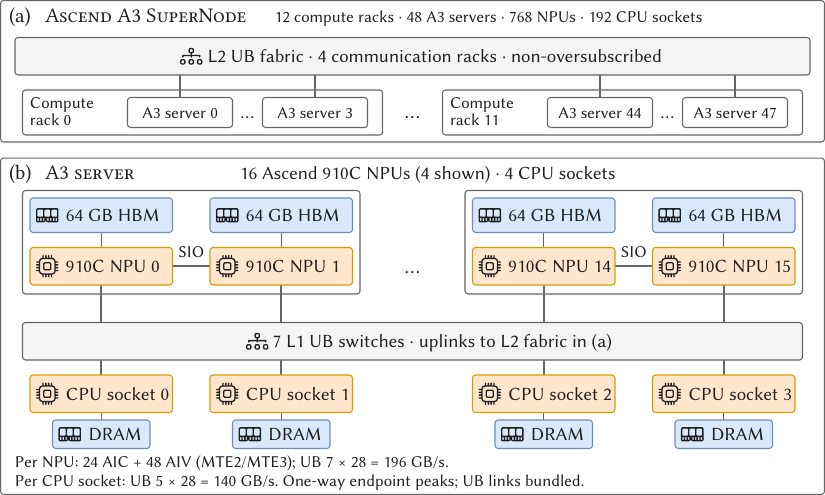}
\caption{Ascend A3 hardware hierarchy and UB interconnect.
(a) The A3 supernode contains 48 A3 servers across 12 compute racks, connected
through the L2 UB fabric in four communication racks.
(b) Each A3 server has seven L1 UB switches, 16 910C NPUs (four shown), and
four CPU sockets. Each NPU connects to HBM, each CPU connects to DRAM, and SIO
connects paired NPUs.
UB links are drawn as bundles; individual switch-plane wiring is omitted.
Bandwidth annotations are one-way endpoint peaks for the reference configuration.}
\label{fig:ascend-supernode}
\end{figure}

Figure~\ref{fig:ascend-supernode} shows these boundaries. Second-level
UB switches connect the 48 A3 servers without oversubscribing the scale-up
fabric~\citep{cloudmatrix}. Here, \emph{cross-server} means crossing A3 servers
within one supernode; it does not mean crossing the separate scale-out network
between supernodes.

Each 910C NPU has 64\,GB of HBM, 24 AI Cube (AIC) cores, 48 AI Vector (AIV)
cores, and up to 1.6\,TB/s of local memory bandwidth. The four CPU sockets of an
A3 server form four NUMA domains. All decoder--service traffic in our evaluation
shares the NPU ingress and egress path of this fabric.
Paired NPUs are connected
by the SIO interconnect~\citep{cloudmatrix}. UB attaches
both NPU memory and CPU DRAM to the scale-up fabric. Placing the indexer on
another server therefore adds no separate low-bandwidth PCIe-to-RNIC path.
The decoder can keep a sparse-KV buffer in HBM, back it with a DRAM-resident
full KV cache, and exchange requests with a service that holds the past index
in service HBM.

\subsection{Global addressing and device-driven access}
UB provides a global physical address space for \emph{fabric-attached} memory.
An NPU's virtual addresses reach this space through its memory mappings and
hardware address translation. The runtime establishes page-table mappings;
the memory management unit translates the virtual address used by a device
kernel into a fabric-attached physical address. This allows byte-addressable
access to mapped remote HBM or CPU DRAM across the supernode. It does not make
every host allocation remotely accessible: private host memory is not the same
as memory configured for fabric access. A global physical address space also
does not mean that different processes use the same virtual addresses.

\sys uses this fabric for small, memory-semantic transfers between the decoder
and the remote indexer service.
At initialization, the runtime allocates communication regions, registers them
with the memory backend, and exchanges the information needed to set up
accessible mappings. Each participant keeps its own mapped address and the
offsets of request, response, and control slots. It keeps those mappings valid
until all graph references and outstanding transfers have drained.
Device kernels running on compute cores then access mapped remote memory
through memory-transfer operations~\citep{ascendcdatacopy}.
These kernels gather current-step inputs, transfer them to the service, and
publish completion without a separate host-issued copy for each field.
The same mechanism returns selected token IDs to the decoder.
This device-driven path is especially useful for the small exchanges repeated
at every model layer.

Global reachability does not give a coherent shared heap.
The runtime orders payload writes before it publishes completion,
checks request identity before consuming a response, and delays buffer reuse
until consumption. These rules keep the graph's addresses stable while the
contents change on each replay. Kernel-driven transfers also use device
resources, so running more of them in parallel can interfere with model
execution.

\subsection{Ascend mapping and concurrent transfer kernels}
The Ascend implementation uses MemFabric's shared-memory mappings to expose the
communication regions to device kernels. A remote location comes from the
mapping's base, the peer's runtime-provided symmetric-region stride, and its slot
offset. The runtime supplies the stride because its virtual layout can reserve
more space per participant than the memory that participant contributes.
The kernel uses this mapping metadata to resolve a peer address, and keeps the
resulting mappings across graph replays.

Ascend's compute and transfer resources make this mapped path programmable.
Each 910C NPU contains the AIC and AIV cores shown in
Figure~\ref{fig:ascend-supernode}. The AIV execution path has on-chip buffering
and two memory transfer engines, MTE2 and MTE3, for moving data between global
memory and the local buffer. Ascend C \texttt{DataCopy} operations express these
transfers~\citep{ascendcdatacopy}. With a valid remote mapping, global memory can
point to a peer's fabric-attached memory. The kernel can therefore gather
decoder fields and write remote slots through the memory path without staging
the numerical payload in a CPU buffer.

The request kernel gives independent fields to several AIV cores so that
their transfer engines can run at the same time. The prototype uses
nine cores to copy nine segments to fixed,
32-byte-aligned offsets. The cores synchronize after their payload transfers;
one chosen core publishes the request marker and owns the response protocol.
The core-local buffer stays part of the transfer path. Memory-semantic
communication therefore avoids a CPU bounce buffer and can remove an
intermediate packed HBM buffer; it does not mean that data skips all buffers or
moves for free.

Synchronization must match the execution resources that take part. The
prototype uses an AIV-only barrier for its communication cores after their
payload transfers. Control transfers are padded to the backend's supported
granularity, which is 32 bytes here. The chosen publishing core then writes the
ready or completion marker after the matching payload is visible.
Appendix~\ref{app:communication} states these ordering rules without
Ascend-specific barriers and alignment, and says which copies are removed.

\subsection{Bandwidth and latency for indexer disaggregation}
Endpoint link rates and sustained memory-transfer bandwidth are different
quantities. In the reference configuration, each NPU has seven
28\,GB/s UB links, totaling 196\,GB/s in one direction; each CPU socket has
five such links, totaling 140\,GB/s. These are aggregate endpoint line rates,
not the bandwidth of every NPU--DRAM pair.

\begin{table}[ht]
\centering
\small
\caption{Memory-access performance from the reference platform characterization.
NPU bandwidth is sustained bandwidth when an NPU reads the indicated
memory. Latency is the reported access latency for that path, not the
decoder--service round-trip time. These measurements are distinct from
endpoint line rates and from \sys's end-to-end results.}
\label{tab:ub-platform-bandwidth}
\begin{tabular}{lrr}
\toprule
Memory accessed over UB & NPU bandwidth (GB/s) & Latency ($\mu$s) \\
\midrule
Other-package HBM, same server & 165 & 0.7 \\
HBM, different server & 160 & 2.1 \\
CPU DRAM, same server & 112 & 0.8 \\
CPU DRAM, different server & 110 & 2.2 \\
\bottomrule
\end{tabular}
\end{table}

Low access latency matters because each layer waits for its selection result
before sparse attention can proceed. Table~\ref{tab:ub-platform-bandwidth}
reports 0.7--2.1\,$\mu$s for remote HBM access, with cross-server bandwidth close
to the in-server path. This supports small exchanges over the scale-up
fabric, although software synchronization and indexer execution add to the
full round trip. \sys keeps scans over the stored index keys inside the service
and sends only current-step inputs and selection results during decode.

\subsection{Striped DRAM backing across CPU sockets}
Larger decoder batches also raise the KV traffic that CPU DRAM must supply.
Host--accelerator bandwidth is a known limit for KV offloading:
ECHO, for example, studies recall over a 64\,GB/s PCIe Gen5 connection
between the host and each GPU~\citep{echo}.
The supernode exposes multiple CPU sockets through UB, but placing a KV pool
behind one socket can still concentrate its reads on that socket's egress.
This matters most for the baseline's shared full-KV pool in DRAM:
tensor-parallel ranks in the same data-parallel group share one DRAM-resident
full KV cache~\citep{vllmsfa}.
Sharing stores one copy while several ranks can still fetch KV at the same
time.

We stripe the pool's physical backing across CPU sockets so KV reads can
use their combined egress bandwidth. We apply the same CPU-socket striping optimization to the DRAM-resident full-KV pools of all four evaluated systems: vLLM, SGLang HiSparse, ECHO, and \sys as a system-level memory-placement optimization rather than an \sys-specific advantage. A stable logical address space maps successive stripes to different sockets;
decoder block handles and captured graph pointers do not change.
The stripes must be fine enough to spread a layer's working set, not just to
balance total allocation across all layers.
This mapping keeps one CPU socket's supply bandwidth from becoming the
bottleneck at large batches.
It does not remove contention at NPU ingress or elsewhere in the shared
fabric.
 \section{Evaluation Details}
\label{app:eval-setup}

\subsection{Setup}
\paragraph{Hardware and models.}
All servers belong to the same \ascendAThree supernode
(Appendix~\ref{app:hardware}). Each A3 server has 16 Ascend 910C NPUs.
The main comparison uses one decoder server and a separate service server.
The scalability experiment uses up to 14 decoder servers with one service
server. DeepSeek-V3.2 INT8 occupies 647~GiB across 61 layers; GLM-5.1 INT4
occupies 391~GiB across 78 layers. Both models use BF16 MLA KV entries;
all systems keep the full KV cache in DRAM.
Like ECHO~\citep{echo}, we separate prefill from decode. Each request
starts generation after its precomputed KV cache and index have been
loaded. \sys keeps the index in service HBM.

\paragraph{Software versions and Ascend ports.}
We run SGLang v0.5.17 and vLLM v0.26.0 with their Ascend backends, using vLLM
Ascend v0.26.0rc1. Our HiSparse baseline uses an Ascend port under internal
development within the SGLang community. We reimplement ECHO's early KV
prefetching and its fusion into the indexer kernels in SGLang's Ascend backend.
All baselines support graph execution.

\paragraph{Baseline decode path.}
During decoding, each baseline offloads the full KV cache to a DRAM
pool and keeps the full index in HBM. At each step, it fills an HBM-resident
sparse-KV buffer from the indexer's top-$k$ selection, reusing cached entries
and fetching only misses.

\paragraph{Baseline settings.}
The main throughput comparison uses \texttt{max-num-seqs=16} for
vLLM~\citep{vllmsfa} and \texttt{host\_to\_device\_ratio=5} for
SGLang HiSparse and ECHO~\citep{hisparse}. The latter sets the ratio of
index-token capacity to sparse-KV-token capacity.
Appendix~\ref{app:partition} varies the static capacity split.
The separate decoder-parallelism sweep uses vLLM
\texttt{max-num-seqs=24} (Appendix~\ref{app:model-parallel}).

\paragraph{Configuration for the model analysis in Section~\ref{sec:motivation}.}
That analysis uses the DeepSeek-V3.2 INT8 model on vLLM Ascend
v0.26.0rc1-a3 with TP8 $\times$ DP2, EP16, ACLGraph, vLLM sparse offloading,
$k=2048$, and a 4096-token sparse-KV buffer per request per layer.
For Figure~\ref{fig:memory}a, we replay each stage for over 20 seconds on two
TP8 replicas and time its graph with device events. Throughput is normalized to
each stage's maximum throughput. Larger batches did not increase throughput.
Hardware checks show that these peaks are limited by compute, bandwidth, or
data movement. The
INT8-quantized indexer and sparse attention use a 64K context; dense attention
uses 32K and 64K contexts with 16 heads per NPU; MoE includes routing and
communication.
Figure~\ref{fig:memory}b samples the same three context-length groups as
Workloads A--C and uses vLLM sparse offloading with \texttt{max-num-seqs=16}.
It replays Poisson arrivals with the same 128-token output cap.
Figure~\ref{fig:memory}c excludes runtime overhead and page rounding; model
weights occupy 45.4~GiB of each NPU's 61.3~GiB HBM, and the remaining 15.9~GiB
is split between the sparse-KV buffer and index.  The capacity bound is an idealized example and the
stage profile is a microbenchmark, so the two show different aspects of the
same deployment rather than one configuration.

\paragraph{Resource accounting.}
Throughput counts completed output tokens per second. For a 16-NPU decoder
and $n$ service NPUs, we multiply \sys throughput by $16/(16+n)$.
Unused NPUs on the service server are excluded from this per-NPU accounting.
The scalability comparison instead counts whole servers, each with 16 NPUs.
At a total of $m$ servers, \sys uses $m-1$ decoder servers and one service
server; vLLM uses all $m$ for decoding.
Mean ITL is the time between consecutive output tokens.
 \section{Sensitivity to Static Capacity Partitions}
\label{app:partition}

Sparse-offloading runtimes split cache HBM between an index and a sparse-KV
staging buffer. The split is fixed before a request arrives, even though the two
pools grow differently with context length and concurrency. This appendix first
shows how vLLM sparse offloading and SGLang HiSparse expose that split. It
then gives the full vLLM sensitivity sweep we used to pick the main-paper
configuration.

\paragraph{vLLM sparse offloading: per-request sparse-KV slots.}
vLLM configures a maximum number of concurrent sequences through
\texttt{max-num-seqs}. Let this value be $C$. The runtime reserves a sparse-KV
buffer for each of the $C$ possible sequences and uses the remaining cache HBM
for index entries. In our configuration, each admitted request can stage 4,096
sparse-KV tokens per layer. If $M$ is the cache-HBM budget, $m_s$ is the
per-request sparse-KV-buffer footprint, and $m_i(L)$ is the index footprint of a
request with context length $L$, a simplified capacity bound is
\begin{equation}
 B_C(L)=\min\!\left(C,\frac{M-Cm_s}{m_i(L)}\right),
 \qquad M\geq C m_s.
 \label{eq:vllm-partition}
\end{equation}
This expression leaves out allocator metadata, page rounding, graph workspaces,
and other runtime allocations. A small $C$ can cap the batch while leaving index
capacity unused. A large $C$ reserves more sparse-KV slots but leaves less HBM
for long-context indexes. We evaluate $C\in\{8,16,24,32\}$.

\paragraph{SGLang HiSparse: index-to-sparse-KV token ratio.}
SGLang HiSparse instead sets index-token capacity as a multiple of sparse-KV
token capacity~\citep{hisparse}. Let $R$ be this
\texttt{host\_to\_device\_ratio}, $T_s$ be the aggregate sparse-KV-token
capacity, and $T_i=R T_s$ be the index-token capacity. If $c_s$ and $c_i$ are
the corresponding bytes per token, the cache-HBM constraint gives
\begin{equation}
 T_s=\frac{M}{c_s+R c_i},\qquad
 B_R(L)=\min\!\left(\frac{T_s}{h},\frac{R T_s}{L}\right),
 \label{eq:partition-ratio}
\end{equation}
where $h$ is the sparse-KV reservation per admitted request. A larger $R$
gives more token capacity to the index and less to sparse-KV staging. So
vLLM fixes the number of per-request staging slots, whereas SGLang fixes a ratio
between two aggregate token capacities. Both are static, and their best
settings depend on the context-length and concurrency distribution.

\begin{figure}[H]
\centering
\includegraphics[width=.98\linewidth]{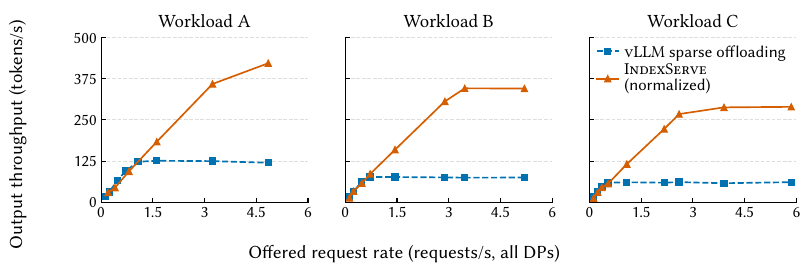}
\caption{DeepSeek-V3.2 INT8, TP8$\times$DP2, with vLLM
\texttt{max-num-seqs=8}. Panels show Workloads A, B, and C. \sys is normalized
by total allocated NPUs; vLLM uses the 16 decoder NPUs.}
\label{fig:static-deepseek-8}
\end{figure}

\begin{figure}[H]
\centering
\includegraphics[width=.98\linewidth]{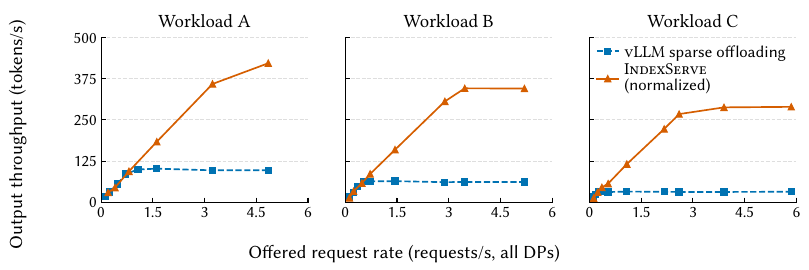}
\caption{DeepSeek-V3.2 INT8, TP8$\times$DP2, with vLLM
\texttt{max-num-seqs=16}. Panels show Workloads A, B, and C; \sys throughput is
resource-normalized.}
\label{fig:static-deepseek-16}
\end{figure}

\begin{figure}[H]
\centering
\includegraphics[width=.98\linewidth]{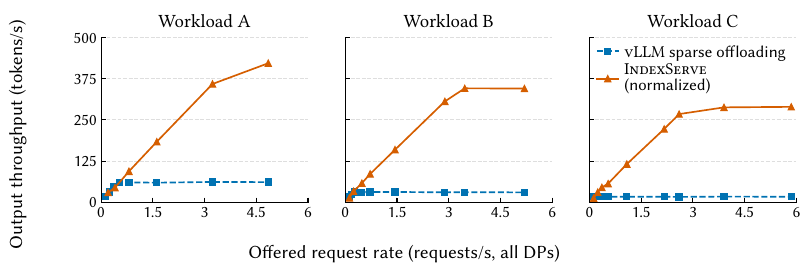}
\caption{DeepSeek-V3.2 INT8, TP8$\times$DP2, with vLLM
\texttt{max-num-seqs=24}. Panels show Workloads A, B, and C; \sys throughput is
resource-normalized.}
\label{fig:static-deepseek-24}
\end{figure}

\begin{figure}[H]
\centering
\includegraphics[width=.98\linewidth]{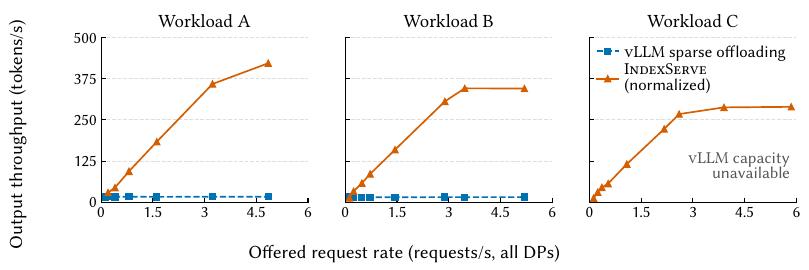}
\caption{DeepSeek-V3.2 INT8, TP8$\times$DP2, with vLLM
\texttt{max-num-seqs=32}. Panels show Workloads A, B, and C. vLLM cannot admit
Workload C; \sys throughput is resource-normalized.}
\label{fig:static-deepseek-32}
\end{figure}

\begin{figure}[H]
\centering
\includegraphics[width=.98\linewidth]{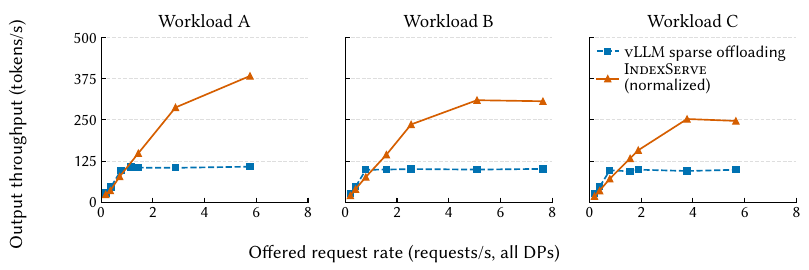}
\caption{GLM-5.1 INT4, TP8$\times$DP2, with vLLM
\texttt{max-num-seqs=8}. Panels show Workloads A, B, and C; \sys throughput is
resource-normalized.}
\label{fig:static-glm-8}
\end{figure}

\begin{figure}[H]
\centering
\includegraphics[width=.98\linewidth]{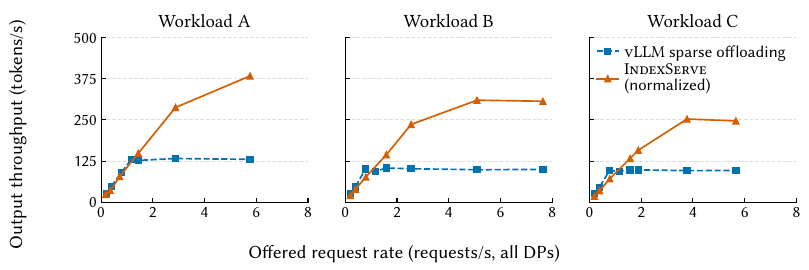}
\caption{GLM-5.1 INT4, TP8$\times$DP2, with vLLM
\texttt{max-num-seqs=16}. Panels show Workloads A, B, and C; \sys throughput is
resource-normalized.}
\label{fig:static-glm-16}
\end{figure}

\begin{figure}[H]
\centering
\includegraphics[width=.98\linewidth]{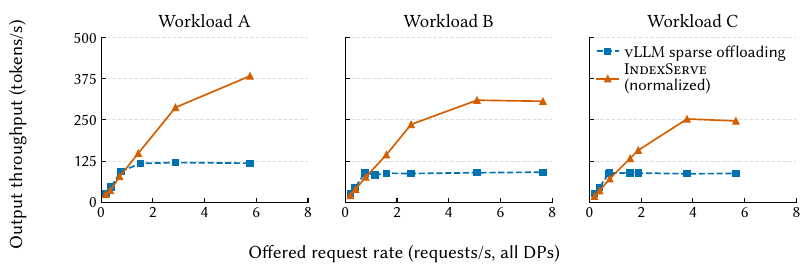}
\caption{GLM-5.1 INT4, TP8$\times$DP2, with vLLM
\texttt{max-num-seqs=24}. Panels show Workloads A, B, and C; \sys throughput is
resource-normalized.}
\label{fig:static-glm-24}
\end{figure}

\begin{figure}[H]
\centering
\includegraphics[width=.98\linewidth]{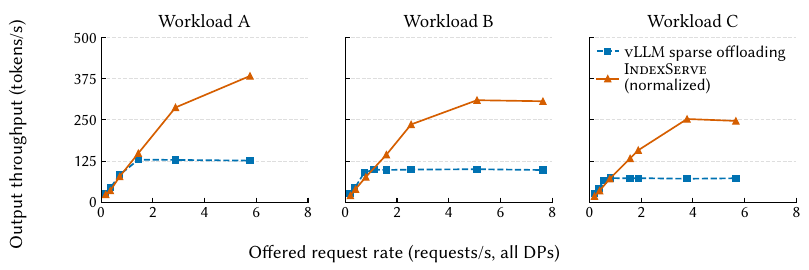}
\caption{GLM-5.1 INT4, TP8$\times$DP2, with vLLM
\texttt{max-num-seqs=32}. Panels show Workloads A, B, and C; \sys throughput is
resource-normalized.}
\label{fig:static-glm-32}
\end{figure}

\paragraph{Methodology.}
Figures~\ref{fig:static-deepseek-8}--\ref{fig:static-glm-32} compare only vLLM
sparse offloading and \sys. Both models use TP8$\times$DP2 on 16 decoder NPUs.
Each figure follows the Workload A/B/C definitions in
Section~\ref{sec:eval-setup}.
The setting in each title applies only to vLLM; \sys has no corresponding static
partition. We normalize each \sys point by $16/(16+n)$, where $n$ is the
service-NPU count for that model, workload, and setting, and label it
\emph{IndexServe (normalized)}. DeepSeek-V3.2 under Workload C with
\texttt{max-num-seqs=32} is capacity-unavailable.

 \section{Sensitivity to Decoder Parallelism}
\label{app:model-parallel}

This appendix reports request-rate sweeps for the DP$\times$TP layouts in
Section~\ref{sec:evaluation}. We use GLM-5.1 INT4 and hold the decoder budget
at 16 NPUs.
The four layouts are DP16$\times$TP1, DP8$\times$TP2, DP4$\times$TP4, and
DP2$\times$TP8. A larger TP therefore leaves fewer independent DP replicas
while the decoder budget stays fixed. This sweep also tests whether \sys's gain
comes only from removing rank-local index replication under TP. The colocated
baseline's replication factor falls with TP and is gone at TP1. A
resource-normalized gain at TP1 therefore cannot come from removing TP
replication. The full curves across the four layouts show that the gain stays
as the replication factor changes. All figures use the
\texttt{max-num-seqs=24} source data and follow the Workload A/B/C definitions
in Section~\ref{sec:eval-setup}. The setting applies to vLLM; \sys has no
corresponding static partition.

The horizontal axis is the total offered request rate across all DP
replicas. The curves therefore show both how many independent replicas a
layout has and how each wider TP group runs a request. We plot only
vLLM sparse offloading and \sys. Each \sys point is labeled
\emph{IndexServe (normalized)} and scaled by $16/(16+n)$ using the service-NPU
count $n$ recorded for that layout and workload.

\begin{figure}[H]
\centering
\includegraphics[width=.98\linewidth]{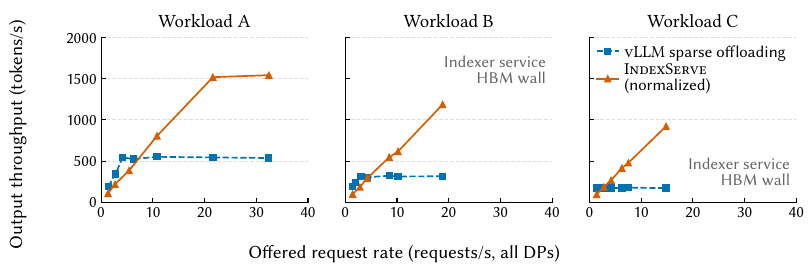}
\caption{GLM-5.1 INT4 request-rate sweeps with TP1$\times$DP16 on 16 decoder
NPUs and vLLM \texttt{max-num-seqs=24}. Panels show Workloads A, B, and C;
Workloads B and C reach the HBM capacity of one A3 server allocated to the
Indexer Service.}
\label{fig:parallel-glm-tp1}
\end{figure}

\begin{figure}[H]
\centering
\includegraphics[width=.98\linewidth]{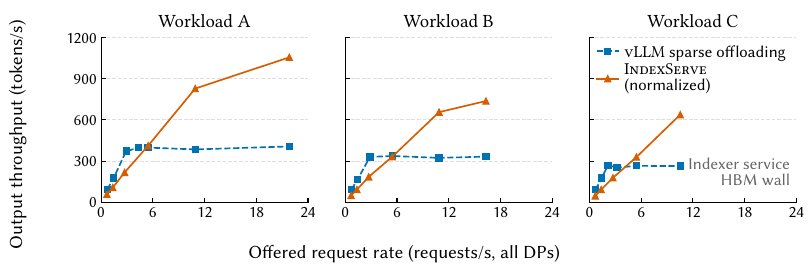}
\caption{GLM-5.1 INT4 request-rate sweeps with TP2$\times$DP8 on 16 decoder
NPUs and vLLM \texttt{max-num-seqs=24}. Panels show Workloads A, B, and C;
Workload C reaches the HBM capacity of one A3 server allocated to the Indexer
Service. }
\label{fig:parallel-glm-tp2}
\end{figure}

\begin{figure}[H]
\centering
\includegraphics[width=.98\linewidth]{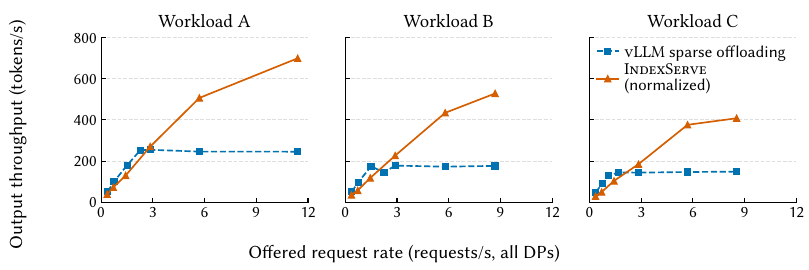}
\caption{GLM-5.1 INT4 request-rate sweeps with TP4$\times$DP4 on 16 decoder
NPUs and vLLM \texttt{max-num-seqs=24}. }
\label{fig:parallel-glm-tp4}
\end{figure}

\begin{figure}[H]
\centering
\includegraphics[width=.98\linewidth]{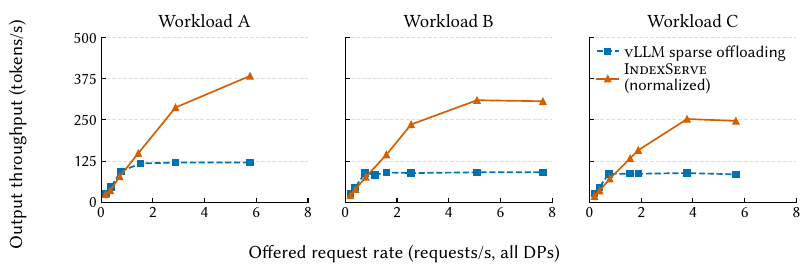}
\caption{GLM-5.1 INT4 request-rate sweeps with TP8$\times$DP2 on 16 decoder
NPUs and vLLM \texttt{max-num-seqs=24}. }
\label{fig:parallel-glm-tp8}
\end{figure}
 \clearpage
\section{Design and Communication Details}
\label{app:design-details}

This appendix expands the service setup, graph execution, memory management,
and scheduling in Section~\ref{sec:design}. It also gives the payload and
metadata behind the fixed-buffer interface in Section~\ref{sec:graph}.
Appendix~\ref{app:hardware} describes the Ascend implementation.

\subsection{Session setup and service boundary}
The proxy assigns a new session to one service accelerator using both free
index capacity and expected scan load. Admission counts index pages,
page tables, graph workspaces, and communication buffers. The service loads
the initial index and publishes its logical handle only after that index and
the decoder's KV backing are ready. This first load sends the full prefix
index. The session keeps the same owner during ordinary decoding. The proxy
sets up this route but does not have to relay each call's tensor data through
CPU memory.

DSA separates model-dependent projections from a scan over the stored index
keys (Equation~\ref{eq:dsa-score}). At layer $\ell$ and token step $t$, the decoder
produces $H_I$ index queries $\mathbf q^I_t\in\mathbb R^{H_I\times d_I}$,
one new index key $\mathbf k^I_t\in\mathbb R^{d_I}$, and scoring coefficients
$\mathbf w^I_t\in\mathbb R^{H_I}$~\citep{deepseekv32}. The decoder keeps these
projections, including the model's positional processing and the chosen
numerical format. The service owns the past index and the operators that
consume these projected values.

A service task covers one session, token step, and indexer-bearing layer.
Tasks may share a batch only when they have the same owner and compatible
scoring operations and numerical formats. For each task, the service appends
the new key and its scale to the correct sequence and layer, scans the
causally visible keys, applies the same top-$k$ rule, and returns the selected
token identities. The scan sees the update so that the current token can be
selected. The full score vector stays on the service. Model weights, the
decoder hidden state, and the stored index keys do not cross the boundary on
each decode call. The decoder uses the returned identities to find
sparse-KV-buffer hits and fetch missing KV before running sparse attention.
Each session retains its own index, valid length, and output when tasks share
a batch.

\subsection{Per-request payload and communication protocol}
\label{app:communication}
DeepSeek-V3.2 specifies $H_I=64$, $d_I=128$, and $k=2048$~\citep{deepseekconfig}.
Table~\ref{tab:dsa-payload} counts the numerical payload for the prototype's
INT8-quantized indexer format. It uses INT8 queries and index keys, FP16 scales
and scoring coefficients, and INT32 selected identities. A query has one scale
per head, while each new index key has one scale. Transport alignment,
graph-bucket padding, and control fields add to the actual transfer.

\begin{table}[ht]
\centering\small
\setlength{\tabcolsep}{5pt}
\caption{Numerical payload for one request at one indexer-bearing layer in the
DeepSeek-V3.2 INT8-quantized indexer format. Descriptor, completion, and optional
miss-plan bytes are additional; the entries do not describe a padded
transport-buffer allocation.}
\label{tab:dsa-payload}
\begin{tabular}{lllr}
\toprule
Direction & Field & Shape and format & Bytes \\
\midrule
Decoder $\to$ service & Index queries & $64\times128$ INT8 & 8,192 \\
 & Query scales & $64$ FP16 & 128 \\
 & Scoring coefficients & $64$ FP16 & 128 \\
 & New index key & $128$ INT8 & 128 \\
 & New-key scale & $1$ FP16 & 2 \\
\cmidrule(l){2-4}
 & Input numerical payload & & 8,578 \\
\midrule
Service $\to$ decoder & Selected token IDs & $2048$ INT32 & 8,192 \\
\bottomrule
\end{tabular}
\end{table}

More generally, for element widths $b_q$, $b_w$, $b_I$, and $b_{\rm id}$ in
bytes, the request and response sizes are
\begin{equation}
\begin{aligned}
 S_{\rm req}&=H_Id_Ib_q+H_Ib_w+d_Ib_I
                 +S_{\rm scales}+S_{\rm descriptor},\\
 S_{\rm resp}&=k b_{\rm id}+S_{\rm result\ metadata}.
\end{aligned}
\label{eq:dsa-wire-size}
\end{equation}
The INT8-indexer numerical exchange totals 16,770 bytes, approximately
16.38 KiB per layer. As a separate format example, BF16 queries alone occupy
16 KiB. With BF16 new keys and FP16 coefficients, the unquantized input payload
is 16,768 bytes; adding INT32 top-2048 IDs gives 24.375 KiB. Each example keeps
its own numerical configuration.

A descriptor travels with the numerical payload. It identifies the model
adapter, sequence and generation, layer, token step, invocation, valid length,
index handle, and response slot. These fields specify which resident state to
update and which decoder may consume the result. A compact miss plan also gives
destination slots and backing addresses and carries a residency version.

For fixed $H_I$, $d_I$, and $k$, this per-call numerical exchange does not grow
with context length $L$, although resident index size and scan work do. Setup
loads all $L$ index keys for each indexer-bearing layer. At 128K tokens, this is
16.25 MiB per layer in the INT8 key/scale format, or 32 MiB in BF16. These
initial transfers are counted separately from decode calls. The selected KV
takes a different path, directly from the DRAM-resident full KV cache to the
sparse-KV buffer. For DeepSeek-V3.2, at $k=2048$ and 1152 bytes per BF16 MLA KV
entry, an all-miss layer needs 2.25 MiB per request on that path. Other model
adapters use Equation~\ref{eq:dsa-wire-size} with their own head counts,
dimensions, and numerical formats.

The protocol separates routing information, per-invocation state, and
completion state. Routing information binds a session to its service owner and
registered communication slots. Invocation metadata gives the sequence
generation, layer, token step, model/format, valid index length, and logical
index handle. Completion metadata echoes the invocation identity and records
status so a decoder can distinguish the current result from an earlier use of
the same slot.

On the request path, the decoder writes the numerical inputs and descriptor
before publishing a ready marker. The service resolves the index handle,
installs any needed page mapping, and appends the new key before scan/select.
On the response path, the service writes selected IDs and response identity
before publishing completion. The decoder's device-side wait checks that the
completion belongs to its current call before releasing the dependent KV
lookup, fetch, and attention operations. Payload visibility precedes marker
visibility on both paths, and marker observation precedes payload consumption.
The slot remains assigned to its call until the response has been consumed.
Identity checks and generation changes prevent a late completion from being
used as a new result.

The backend exposes registration, mapped address exchange, request publication,
completion observation, and deregistration after quiescence. A registration
record binds a buffer to its owner, access permissions, size, and lifetime.
Model code uses logical handles and stable buffers, and the backend resolves
them to reachable addresses. A direct-pack kernel writes input fields to their
assigned offsets in the remote slot, removing a separate local packed-request
buffer. Block tables are versioned and updated when mappings change instead of
being copied in full every layer. Zero-copy avoids CPU payload staging and
needless intermediate buffer copies. The service may gather rows into the
contiguous local layout used by its scan operator.

An owner submits one request only after the inputs match across TP ranks. If the
model shards query projections, an assembly step runs before the service call.
Returned IDs go to the ranks that use the same selection. Data and completion
markers use separate ownership rules so consumers cannot overwrite an unread
result.

\subsection{Graph replay and stable addressing}
The decoder captures submission, completion observation, and result use with
fixed request and response buffer addresses. Each replay updates the current
inputs, index handle, and invocation identity in those buffers. A device-side
wait releases KV lookup only after the matching result arrives. A change in
decoder batch configuration selects the corresponding captured graph.

The service captures a separate graph for each supported execution bucket.
An execution bucket fixes the batch shape and the addresses of inputs, outputs,
and workspaces. Before replay, the scheduler fills row descriptors with the
chosen tasks, their index handles, and valid lengths. Unused rows are masked.
Changing row membership or index length changes runtime data rather than the
graph's buffer bindings.

The communication backend maps a registered region into an address that a
device can reach across the fabric. Inside the service, PagedIndex maps each
logical index position to its HBM page. The communication region stays pinned
across replays, while the index's page-table entries grow as the sequence
appends tokens. The page-table directory also has a stable address and enough
metadata capacity for the admitted configuration. Adding an index page updates
an entry before the consuming graph runs without replacing the directory or
moving existing keys. This indirection lets batch membership change and indexes
grow under graph replay.

\subsection{PagedIndex allocation and capacity}
PagedIndex maintains a page table for each session at each indexer-bearing
layer. Let $P$ be the number of keys per page and $s_I$ the bytes per key,
including any scale. Token position $t$ uses logical page $\lfloor t/P\rfloor$
and offset $t\bmod P$. Physical pages need not be adjacent. Scan/select
resolves them through the page table, while the decoder keeps the same logical
handle.

For a prefix of length $L_r$, the initial load allocates
$\lceil L_r/P\rceil$ pages per layer. New keys fill the last page. When it is
full, the allocator takes a free page and publishes its mapping before the
append runs. Earlier keys stay in place. After a session ends, its pages return
to the pool after outstanding reads and writes finish.

For $n_I$ equally sized indexer-bearing layers, the allocated index payload is
\begin{equation}
 M_{\rm index}=\sum_r n_I\left\lceil L_r/P\right\rceil P s_I.
\end{equation}
With $N$ nonempty sessions, unused space in the final pages is less than
$N n_I P s_I$. This bound covers payload pages only; admission also counts
page tables, alignment, graph buffers, workspaces, and scan load. Sessions
share free pages as they grow.

\subsection{Batch sizing and dispatch}
Each owner organizes ready tasks by the valid history length they will scan.
These length groups guide dispatch priority but are not hard batch boundaries.
Compatible tasks from different groups may share a batch, and a growing
session can change groups without moving its stored index.

Offline profiles record execution time $T_c(b)$ for each length group $c$ and
supported batch size $b$. The profiled path includes input preparation, key
append, scan/select, and result preparation. Its rate is $b/T_c(b)$. At each
dispatch, the scheduler considers currently ready tasks and selects a supported
batch size according to the latency objective and current load. A stricter
latency objective favors smaller batches. A throughput-oriented objective or
higher load favors larger batches while avoiding sizes with little additional
throughput. Mixed-length batches are allowed when their predicted execution
time meets the objective. The scheduler does not wait to fill a preferred
batch and masks unused rows in the selected bucket.

A dispatch opportunity lets one eligible batch run. Short-index groups get
more recurring opportunities, reducing the number of long batches a short task
waits behind. Every active long-index group retains recurring opportunities,
and tasks within each group run oldest first. Another ready group can use an
empty group's opportunity. Batch execution budgets limit how long one dispatch
holds the device. With bounded execution times, finitely many older admitted
tasks, and recurring opportunities, long tasks continue to make progress. New
sessions wait when index memory or scan capacity is exhausted.
 \clearpage
\section{Portability to NVIDIA Systems}
\label{app:nvidia-portability}

We implement and evaluate \sys only on Ascend A3. This appendix analyzes the requirements to port \sys's designs to NVIDIA GB200 NVL72 supernode/B200 platforms~\citep{gb200nvl72,blackwellguide}.

\begin{itemize}
  \item \textbf{Service, paging, and scheduling.} Indexer disaggregation,
  PagedIndex's logical page table, admission, and length-aware batching are
  software policies. CUDA kernels can implement append, scan/select, and
  gather/scatter. CUDA virtual-memory management can reserve stable virtual
  addresses and map physical chunks on demand, although PagedIndex can also
  retain its explicit device page table~\citep{cudavmm,vllm}.

  \item \textbf{Graph replay.} CUDA Graphs provide capture and replay with
  fixed topology and buffers, matching our per-batch execution buckets;
  runtime row descriptors carry changing sessions and lengths. NCCL
  collective and point-to-point operations are also
  graph-capturable~\citep{cudagraphs,ncclcudagraphs}.

  \item \textbf{NVLink scale-up.} GB200 NVL72 forms a multi-node NVLink
  domain~\cite{mnnvlguide}. IMEX and CUDA fabric-memory handles export and map GPU memory across
  OS domains, providing the registered remote buffers needed by our request
  slots~\citep{imex,cudavmm}. Device kernels can then move payloads and publish
  completion through an NVIDIA backend rather than the A3 Unified Bus backend.

  \item \textbf{Clusters without a rack-scale NVLink domain.} NVSHMEM exposes
  device-side one-sided put/get, atomics, memory ordering, and point-to-point
  synchronization. Its InfiniBand GPUDirect Async (IBGDA) transport implements
  the InfiniBand control and data paths on the GPU, avoiding a CPU proxy on the
  critical path~\citep{nvshmemapi,nvshmemibgda}. Registered symmetric buffers,
  put-with-signal (or put plus an ordered signal), and device-side waits can
  implement the same ownership, generation, and completion protocol over RDMA.
\end{itemize}

\paragraph{Portability boundary.}
The key mechanisms of \sys do not depend on Ascend's Unified Bus. NVIDIA platforms provide the required capabilities, including remote memory access, device-initiated communication, synchronization, and graph replay. Porting \sys therefore mainly requires adapting the communication backend and kernels, validating memory ordering and completion, and tuning page and batch sizes for the target hardware. While the performance may differ across interconnects, particularly between NVLink and InfiniBand, the same architecture and execution protocol can be implemented on NVIDIA systems.
 \end{document}